\documentstyle[epsfig]{aipproc}

\begin{document}

\title{The Pulsar in GRO J1744$-$28\\ at Low Flux}
 
\author{Michael J.\ Stark$^{*\dagger}$, A.\ M.\ Ahearn$^{*}$,
L.\ J.\ Duva$^*$, and K.\ Jahoda$^{\dagger}$}
\address{$^*$University of Maryland, College Park, MD  20742\\
$^{\dagger}$Laboratory for High Energy Astrophysics, NASA / GSFC, Greenbelt,
	    MD 20771}

\maketitle

\begin{abstract}
The Bursting Pulsar, GRO~J1744$-$28, has been observed with the {\em Rossi
X-ray Timing Explorer\/} ({\em RXTE\/}) regularly since the launch of the
satellite in December 1995 \cite{giles}. This accretion driven pulsar has
undergone two outbursts which have been accompanied by significant spin-up
of the pulsar. Between these two outbursts, the pulsar was detectable
during each observation down to fluxes perhaps one thousandth of the
persistent flux at the peak of the first outburst. We observe evidence of a
change in the sign of the frequency derivative during the faintest period
but the relationship between the observed luminosity and the frequency
derivative is complex.
\end{abstract}

\section*{Introduction}

The instruments aboard {\em RXTE\/} are well suited to the study of
GRO~J1744$-$28.  The large area and high throughput of the {\em RXTE\/}
Proportional Counter Array (PCA) have allowed the study of this source over
a wide dynamic range in luminosity.  The source has already been invoked as
a test of the standard theory of accretion onto magnetized neutron stars
\cite{ghoshlamb}.  Data from the Burst and Transient Source Experiment on
the {\em Compton Gamma Ray Observatory\/} have shown that there is
relatively good agreement between the theoretical dependence of the
accretion torque on the source luminosity and observation when the
source is bright \cite{finger1,finger2}.  This theory also predicts that at
sufficiently low source luminosity the accretion torque should become
negative.  It has been argued elsewhere \cite{cui} that GRO~J1744$-$28
has undergone the transition to negative accretion torque during the period
studied here.  This argument is based on observational parameters other
than changes in the spin period, however.  The magnetic field strength of
the neutron star determines the theoretical form of the dependence of the
accretion torque on the source luminosity.  Knowing the luminosity at which
the transition from positive to negative accretion torque occurs allows a
determination of the magnetic field strength.

\section*{Analysis}

We have searched for evidence of negative accretion torque in the pulsar
signal from GRO~J1744$-$28.  In addition, we have determined the source's
luminosity and present here the magnetic field strength implied by the
luminosity when the accretion torque changes sign.

\subsection*{Torque}

Observations were made of GRO~J1744$-$28 every one to two days during 1996.
Sinusoidal 2.1 Hz pulsation was detectable during each of these
observations.  The period which best describes the pulsation in each
observation was determined by epoch-folding and $\chi^2$ maximization.
Accounting for the binary motion of the source\cite{finger1}, an arrival
time was also calculated for each observation.  Since the pulsar
light-curve is nearly sinusoidal, the arrival time of the pulsation is
taken to be the zero time of the cosine function which best describes the
folded data.

\begin{figure}
\centerline{\epsfig{file=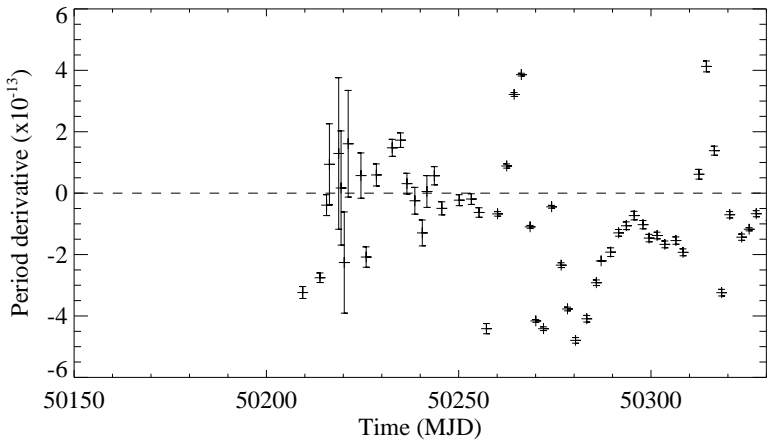,width=4.5in}}
\caption{GRO~J1744$-$28 2.1 Hz pulsar period derivative.  The derivative at
each observation is determined from a fit to the arrival times for five
consecutive observations so neighboring points are not
independent.}\label{pdotfig}
\end{figure}

Combining the best-fit period and arrival times of consecutive
observations, the number of pulses between successive observations can be
counted unambiguously.  The pulse numbers and arrival times can be fit to a
quadratic expression where the coefficient of the quadratic term is
$P_0\dot{P}/2$, half the product of the initial period and period
derivative of the pulsation.  The data from all the observations can be fit
to this expression at once under the assumption that $\dot P$ is constant.
Since we expect that $\dot P$ is changing, we have fit the data in such a
way that a relatively independent $\dot P$ is found for each day.  The
arrival times for five consecutive observations were fit to determine
$\dot{P}$ for the middle observation.  In this way, the $\dot{P}$ for each
observation is dependent only on nearby observations.  The result of this
analysis is shown in Figure \ref{pdotfig}.

\subsection*{Flux}

The data taken on 26 June 1996 using the {\em RXTE\/} PCA are divisible
into two categories: quiet stretches in which the flux was relatively
constant and no pulsation is detected; and {\em rumbly\/} stretches in
which the flux is highly variable and 2.1 Hz pulsation is detected.  For
this analysis, the quiet stretches are taken to be emission from background
sources absent any emission from GRO~J1744$-$28.  The {\em rumbly\/}
stretches are taken to be emission from both background sources and
GRO~J1744$-$28.  Under these assumptions, source and background
contamination can be unambiguously separated.  The energy spectra of the
separated components of these data are fit by typical power-law pulsar
spectra with indices of $-1.52\pm 0.02$ and 
$-2.39\pm 0.01$ respectively.

\begin{figure}
\centerline{\epsfig{file=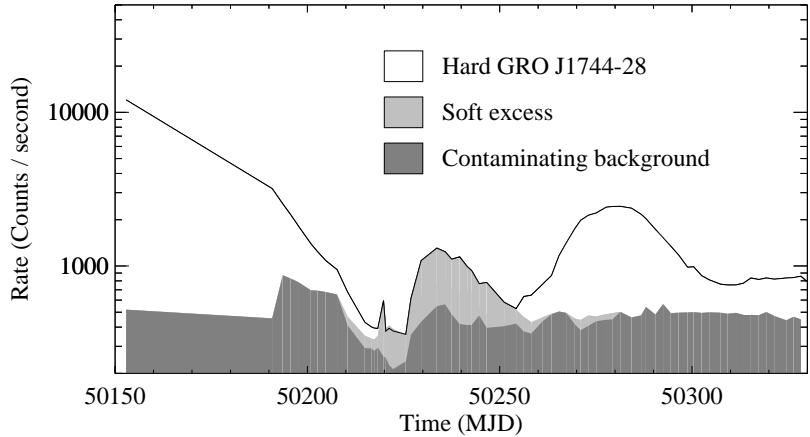,width=4.5in}}
\caption{The counts detected in the PCA from the direction of
GRO~J1744$-$28 for the same time period covered by Figure \ref{pdotfig}.
The proportion of counts in each category is determined by a spectral fit
involving a spectrum typical of GRO J1744-28, a spectrum describing
contaminating background and an extra soft spectral
component.}\label{fluxfig}
\end{figure}

Only the data on 26 June 1996 were unambiguously separable in this way so
data on other days are separated into background and source components by
assuming that the spectra of these two components are invariant.  Each
day's data are then fit by a sum of the spectra described above where only
the total flux from each component is allowed to vary.  The results of
applying this analysis are shown in Figure \ref{fluxfig}.
During outburst,
this analysis suggests a background source which varies slowly, independent
of the large changes in the flux from GRO~J1744$-$28.  During the quietest
period, decomposing the data into two sources is confounded by additional
soft emission from GRO~J1744$-$28 which is incompatible with the assumption
of spectral invariance.  There is an increase in the amplitude of the 2.1
Hz pulsation which is correlated with the appearance of this soft emission
which suggests that this emission is from GRO~J1744$-$28 and, thus, that
its spectrum varies.  For this analysis, we have added an additional soft
component to reflect the spectral change in the source emission.  The
detected counts from this component is shown in Figure \ref{fluxfig}
with the two components which describe the 26 June data.

\section*{Results}

Figure \ref{pdotfig} shows a period of spin-down between MJD 50217 and MJD
50236.  Discounting MJD 50220 when the flux was high and the
period derivative, $\dot P$, was negative for a single observation and
confining analysis to the period when the detected flux was the lowest, the
average $\dot P$ for this period was $(6.8\pm 12.3)\times 10^{-14}$.  This
corresponds to a frequency derivative of $(-3.1\pm7.9)\times 10^{-13}$
s$^{-2}$.

Taking the average flux from GRO~J1744$-$28 between MJD 50231 and MJD 50236
and a distance of 8 kpc, the 2--60 keV luminosity of the source is
$6.7\times 10^{35}$ erg/s.  We can derive the magnetic field implied by
this luminosity by equating the radius of the orbit with a Keplerian
frequency of 2.1 Hz to the radius of the magnetosphere given by Lamb et
al. \cite{lamb}.  This yields the expression
\begin{displaymath}
{B\over 10^{12}{\rm G}}=0.45P^{7/6}\left({M\over 1.4M_\odot}\right)^{1/3}
\left({R\over 10^6{\rm cm}}\right)^{-5/2} \left({L\over 10^{37}{\rm erg / s}}
\right)^{1/14}
\end{displaymath}
where $B$ is the polar surface magnetic field. For typical neutron star
mass and radius, the magnetic field implied by the luminosity we observe is
$1.5\times 10^{11}$ Gauss.  This value is consistent with upper-limits
based on other observational considerations \cite{finger1,bildsten} and
lower than the value derived by Cui\cite{cui} who assumed all counts above
background to be from GRO~J1744$-$28.


\begin{references}

\bibitem{giles} Giles, A.\ B., Swank, J.\ H., Jahoda, K., Zhang, W.,
Strohmayer, T., Stark, M.\ J., \& Morgan, E.\ H. 1996, ApJ, 469, L25

\bibitem{ghoshlamb}Ghosh, P.\ \& Lamb, F.\ K.  1979, ApJ, 234, 296

\bibitem{finger1} Finger, M.\ H., Koh, D.\ T., Nelson, R.\ W., Prince, T.\
A., Vaughan, B.\ A., \& Wilson, R.\ B. 1996, Nature, 381, 291

\bibitem{finger2} Finger, M.\ H., Wilson, R.\ B., \& Chakrabarty, D.  1996,
A\&AS, 120, C209

\bibitem{cui}
Cui, W. 1997, ApJ, 482, L163, astro-ph/9704084

\bibitem{lamb} Lamb, F.\ K., Pethick, C.\ J., \& Pines, D. 1973, ApJ, 184,
271

\bibitem{bildsten} Bildsten, L.\ \& Brown, E.\ F. 1997, ApJ, 477, 897,
astro-ph/9609155

\end{references}
\end{document}